\documentstyle[12pt]{book}
\advance\textheight by \topskip
\catcode`\@=11 
\def\section{\@startsection {section}{1}{\z@}{10mm plus 2.85mm minus
 .6mm}{5mm plus .4mm}{\normalsize\bf}}
\def\subsection{\@startsection{subsection}{2}{\z@}{10mm plus 2.85mm minus
 .6mm}{5mm plus .4mm}{\normalsize\it}}
\def\subsubsection{\@startsection{subsubsection}{3}{\z@}{10mm plus 2.85mm
 minus .6mm}{5mm plus .4mm}{\normalsize\it}}
\def\paragraph{\@startsection
 {paragraph}{4}{\z@}{3.25ex plus 1ex minus .2ex}{-1em}{\normalsize\it}}
\def\subparagraph{\@startsection
 {subparagraph}{4}{\parindent}{3.25ex plus 1ex minus
 .2ex}{-1em}{\normalsize\it}}
\long\def\@makecaption#1#2{
 \vskip 10pt
 \setbox\@tempboxa\hbox{\footnotesize #1: #2}
 \ifdim \wd\@tempboxa >\hsize {\baselineskip 12pt \footnotesize #1: #2\par}
 \else \hbox to\hsize{\hfil\box\@tempboxa\hfil}
 \fi}
\def\thebibliography#1{\section*{References}
 \list{\arabic{enumi}.}%
  {\settowidth\labelwidth{#1.}\leftmargin\labelwidth
  \advance\leftmargin\labelsep
  \rightmargin 0pt
  \itemsep 0pt
  \parsep 0pt
  \usecounter{enumi}}
 \def\newblock{\hskip .11em plus .33em minus -.07em}
 \sloppy\clubpenalty4000\widowpenalty4000
 \sfcode`\.=1000\relax}

\mark{{}{}}
\def\ps@myheadings{\let\@mkboth\@gobbletwo
\def\@oddhead{\hbox{}\footnotesize\sl\rightmark\hfil\rm\thepage}
\def\@oddfoot{}
\def\@evenhead{\footnotesize\rm\thepage\hfil\sl\leftmark\hbox{}}
\def\@evenfoot{}
\def\sectionmark##1{}
\def\subsectionmark##1{}}
\renewcommand{\title}[1]{\newcommand{\titleofthearticle}{#1}}
\newcommand{\titlehead}[1]{\newcommand{\runninghead}{#1}}
\renewcommand{\author}[1]{\newcommand{\authorofthearticle}{#1}}
\newcommand{\address}[1]{\newcommand{\addressoftheauthor}{#1}}
\renewcommand{\maketitle}[1]{%
\thispagestyle{plain}
\markboth{\authorofthearticle}{\runninghead}
\setcounter{equation}{0}
\setcounter{figure}{0}
\setcounter{table}{0}
\setcounter{section}{0}
\begin{center}\begin{normalsize}
\begin{bf}
\titleofthearticle\\
\end{bf}
\bigskip
\authorofthearticle\\
\addressoftheauthor
\end{normalsize}\end{center}
\vspace*{14mm}
}

\renewcommand{\thesubsubsection}%
{\arabic{section}.\arabic{subsection}.\arabic{subsubsection}}

\def\abstract{\small
\begin{center}
{\bf Abstract\vspace{-.5em}\vspace{0pt}}
\end{center}
\quotation}

\newcommand{\HH}{{\cal H}}
\newcommand{\goesto}{\rightarrow}
\newcommand{\ident}{\equiv}

\title{Monte Carlo Renormalization Group Study of the d=1 XXZ Model%
   \footnote{\begin{tabular}[t]{l}
       To appear in
      {\em Quantum Monte Carlo Methods in Condensed Matter Physics},
      ed.\ M. Suzuki, \\
       World Scientific, 1993
             \end{tabular}}
}
\titlehead{MCRG for d=1 XXZ Model}
\author{M.~A.~Novotny$^1$ and H.~G.~Evertz$^{1,2}$}
\address{
\vskip.5ex
$^1$ Supercomputer Computations Research Institute,\\
Florida State University,\\
Tallahassee, FL ~32306-4052 ~USA\\
\vskip 0.2 truecm
$^2$ Department of Physics and Astronomy\\
and Center for Simulational Physics \\
University of Georgia, \\
Athens, GA ~30602 ~USA\\
\vskip 0.2 truecm
\begin{tabular}[t]{ll}
  {\em email:}&novotny@scri.fsu.edu \\
              &evertz@scri.fsu.edu
\end{tabular}}

\begin{document}

\mbox{\ } \vskip-3.5cm\hfill
            \begin{tabular}[t]{l}
                     \rule{0ex}{1ex}FSU-SCRI-93-116   \\[.3ex]
                                    cond-mat/9309047  \\[.3ex]
                     \rule{0ex}{1ex}September 1993
            \end{tabular}
\vskip1.3cm
\maketitle

\begin{abstract}
We report current progress on the synthesis of methods to
alleviate two
major difficulties in implementing a Monte Carlo Renormalization Group (MCRG)
for quantum systems.  In particular, we have utilized the loop-algorithm
to reduce critical slowing down, and we have
implemented an MCRG method in which
the symmetries of the classical equivalent model need not be fully
understood, since the Renormalization Group is given by the Monte Carlo
simulation.  We report preliminary results obtained when the resulting
MCRG method is applied to the $d$$=$$1$ XXZ model.
Our results are encouraging.  However, since this model has a
Kosterlitz-Thouless transition, it does not yet provide a full test of
our MCRG method.
\end{abstract}

\markboth{M.~A.~Novotny and H.~G.~Evertz}{\runninghead}
\section{Introduction}

Although the Monte Carlo method introduced by Metropolis
{\em et al.}\cite{re:Metrop} has been a useful tool in the study of the
critical behavior of classical systems, its application to the study of
the critical behavior of quantum systems requires that many difficulties
be circumvented.  Some of these difficulties are already
present in the study of
classical systems, while others are present only in the study of quantum
systems.  In this paper we utilize present knowledge
in overcoming the difficulties in such studies, and we apply the method to
the study of the one-dimensional XXZ quantum spin model.

The first difficulty to be surmounted in the study of quantum critical
behavior was the problem of how to apply the Monte Carlo method to
quantum systems.  This difficulty was circumvented when
Suzuki\cite{re:Suzuki1,re:Suzuki2,re:Suzuki3}
proposed to use the formula of Trotter\cite{re:Trotter}
to map any $d$-dimensional quantum system onto a
$d$$+$$1$-dimensional classical equivalent.  Convergence properties
of the Suzuki-Trotter transformation\cite{re:Suzuki4} and of higher-order
decompositions have been described\cite{re:Suzuki5}.
Although this decomposition is applicable in general, it can lead to
other difficulties not encountered in the study of most classical systems.

One of the problems often introduced by the application of the
Suzuki-Trotter transformation is the minus-sign problem.  The
minus-sign problem occurs when the `Boltzmann weights' of the classical
equivalent of a quantum problem are not all positive.  In the
Monte Carlo method Boltzmann weights are interpreted as probabilities,
and the
minus-sign problem to date is a serious one and
can only be decreased.  This is done
by performing a simulation with any chosen probability distribution, and
reweighting to the Boltzmann weights of the classical
equivalent\cite{re:RW1,re:Novotny1,re:RW2,re:RW3,re:RW4,re:RW5}.
Unfortunately, this moves the difficulty from one of principle to one of
acquiring adequate statistics. In general the
desired quantities are ratios of two numbers, each of  which come from
the difference between two large numbers
with statistical errors.   The minus-sign problem is therefore still largely
unsolved.  It is interesting to note that it is possible to make some
classical statistical mechanical models have a minus-sign problem.
For example, the simulation of the $d$$=$$2$ Ising ferromagnet has a minus-sign
problem in a certain representation\cite{re:Novotny1,re:MunNov}.

A difficulty which occurs
in the study of both classical and quantum critical behavior
is the critical slowing down that occurs in the standard Monte Carlo method
near the critical point.  The difficulty is that standard Monte Carlo
algorithms employ a {\it local\/} update procedure, and consequently
`information' diffuses through the lattice slowly in a random walk fashion.
This difficulty was first mastered for the ferromagnetic
Potts model by Swendsen and Wang\cite{re:SwenWang} where the
important realization was to perform {\it non-local\/} updates on
clusters closely related to the critical clusters of the model.
Similar procedures have been devised for a number of
systems and have recently been reviewed \cite{re:CSRev1,re:CSRev2}.
The classical equivalent obtained from
the Suzuki-Trotter transformation often maps onto a {\em vertex} model,
and it is only
recently that algorithms to alleviate critical slowing down for
vertex models have been
devised \cite{re:CSD_V1,re:CSD_V2,re:CSD_V4}.

In the study of phase transitions universal quantities, such as critical
exponents, are desired.  However, a phase transition occurs only in the
thermodynamic limit --- whereas Monte Carlo simulations by necessity
can  be done only on finite lattices.
The best ways to obtain estimates of the
critical exponents from Monte Carlo studies are to use
finite-size scaling methods (for reviews see \cite{re:FSS1,re:FSS2})
or renormalization group (RG) methods\cite{re:WilFis72,re:Wilson1}.
One marriage of Monte Carlo and RG methods was given by
Swendsen\cite{re:RHS1,re:RHS2,re:RHS3}.  In Swendsen's implementation
of the Monte Carlo Renormalization Group (MCRG) one calculates only
correlation functions (not coupling constants).  To
successfully apply this MCRG method, it is important that the RG chosen
preserves the symmetries of the Hamiltonian\cite{re:NovLanSwen}.
However, this presents a problem for quantum MCRG calculations since
typically after the application of the Suzuki-Trotter formula one does
not generally know the underlying `Hamiltonian' of the $d$$+$$1$-dimensional
classical equivalent. If you do not know the `Hamiltonian', how can
you know its symmetries?
In particular, if the minus-sign problem is present, the concept of a
`Hamiltonian' on the classical equivalent is not well defined.  If the
classical equivalent is a vertex model, the constraint due to the
continuity of vertex loops is difficult to preserve in an RG treatment.
This is the difficulty that has delayed progress
on the application of MCRG procedures to quantum systems after the initial
success of quantum MCRG methods to
the transverse Ising model\cite{re:KolbPRL,re:NovLan,re:KolbSuzuki},
for which the underlying classical equivalent Hamiltonian is known and has
no global constraints.
For a review of quantum MCRG see Sec.~8 of Ref.\ \cite{re:Suzuki86}.
Similar difficulties have been encountered in applying Wilson's idea of
real-space blocking\cite{re:Wilson1} to exact-diagonalization studies,
which has only recently been surmounted\cite{re:White1,re:White2,re:MANO1}.

A method to overcome the difficulty of deciding what RG procedure
to use in quantum studies
was recently reported by M\"unger and Novotny\cite{re:MunNov}.
The idea is to modify a momentum-space MCRG developed by
Swendsen\cite{re:RHS3}, but to let the Monte Carlo simulation itself
decide which RG to use.  In momentum space this leads to the RG
procedure of systematically discarding the least important degrees of
freedom as one renormalizes the lattice.  This momentum-space MCRG
procedure was successfully applied to the square lattice
$q$-state Potts model with $q$ not restricted to integer values.
Thus the MCRG in Ref.\ \cite{re:MunNov} was performed on
a staggered 6-vertex model which is obtained from
a mapping of the $q$-state Potts model.

In this paper we  apply this MCRG for the first time to
a quantum system --- the 6-vertex model, which is the classical equivalent
of the $d$$=$$1$ spin-$1\over 2$ quantum XXZ model.
The XXZ Hamiltonian is
\begin{equation} \label{eq:XXZHam}
\HH = + \sum_{i=1}^N \left(\sigma^x_i\sigma^x_{i+1}
+ \sigma^y_i\sigma^y_{i+1} +\lambda\sigma^z_i\sigma^z_{i+1}
\right),
\end{equation}
with partition function $Z \!=\! \exp{(-\beta \HH)}$.
Periodic boundary conditions on the chain of $N$ spins are used.
Notice that $\HH$ is invariant under
$(\beta \goesto -\beta$, $\lambda \goesto -\lambda)$
by rotating every second spin through an angle of $\pi$,
and is thus equivalent to the Hamiltonian
$\HH \!=\! - \sum_{i=1}^N \left(\sigma^x_i\sigma^x_{i+1}
+ \sigma^y_i\sigma^y_{i+1}-\lambda\sigma^z_i\sigma^z_{i+1} \right)$.

The case $\lambda$$=$$1$ corresponds to the
isotropic Heisenberg antiferromagnet,
while the case $\lambda$$=$$0$ corresponds to the quantum XY chain.

We have chosen this model since both the loop-algorithm
of Ref.\ \cite{re:CSD_V1,re:CSD_V2,re:CSD_V4},
and the MCRG program of Ref.\ \cite{re:MunNov} were written to study
the 6-vertex model, which is the classical equivalent of the
Hamiltonian in Eq.~(\ref{eq:XXZHam}).

\vfill 
\section{Monte Carlo Method}

It is well known \cite{re:Suzuki1,re:Suzuki4,re:BarmaShastry,re:MarcuWiesler},
that the $d$$=$$1$ XXZ chain
can be mapped to a $(1$$+$$1)$ dimensional classical spin system
by Suzuki's method.
The main steps are a breakup of the Hamiltonian for an $N$-spin chain
into pieces living on
even and odd bonds, an application of the Trotter-Suzuki formula,
and an insertion of complete sets of intermediate states.
The result is a system of classical spins $s_{i,j}=\pm 1$
on a periodic $N\times M$ chessboard lattice
with the Euclidean time (Trotter) direction having extension $M$,
and four-spin interactions on all ``black'' plaquettes of the
chessboard \cite{re:Suzuki1,re:Suzuki4,re:BarmaShastry,re:MarcuWiesler}.
The mapping becomes exact as $M\goesto\infty$.

The four spins $(s_{i,j};s_{i+1,j};s_{i,j+1};s_{i+1,j+1})$
bordering an interacting plaquette obey the continuity constraint
\begin{equation} \label{continuity}
 s_{i,j} + s_{i+1,j} = s_{i,j+1} + s_{i+1,j+1}.
\end{equation}
The {\em world lines} connecting sites with $s_{i,j}=1$ are thus continuous.

This classical spin system is a vertex model:
vertices are located at the center of interacting plaquettes;
they are connected by lines. Each such line touches a lattice
site $(i,j)$.
If $s_{i,j}=$$+1$ $(-1)$, then we
place an arrow that points upwards (downwards) in the Trotter direction
on this line.
Note that the resulting lattice of arrows and vertices has lines
tilted 45 degrees w.r.t.\ the chessboard lattice.

\begin{figure}[tbp]
\vspace*{.6cm}
\center{\em (See \cite{re:BaxterBook}, fig.\ 8.2)}
\vspace*{.6cm}
\caption{The 6 vertices which are allowed in the model are shown.
For the XXZ model the weights are $w_5$$=$$w_6$$=$$1$ and
$w_1$$=$$w_2$$=$$w_3$$=$$w_4$.
\label{fig1}}
\end{figure}

The constraint of Eq.~(\ref{continuity}) is
that of the 6-vertex model: each vertex has two arrows pointing
into the vertex and two pointing away from the vertex.
The six possible vertex configurations are shown in Fig.~\ref{fig1}. 
Their weights $w_i$ in the partition function are the weights of
the interacting plaquettes, namely \cite{re:MarcuWiesler}
\begin{equation}
\begin{array}{lllllll}
a&=&w_1&=&w_2&=& e^{-\lambda \hat\beta},  \\
b&=&w_3&=&w_4&=& e^{\lambda \hat\beta} \sinh 2\hat\beta, \\
c&=&w_5&=&w_6&=& e^{\lambda \hat\beta} \cosh 2\hat\beta,  \\
\end{array}
\end{equation}
with $\hat\beta=\beta/(2M)$.
The 6-vertex model is exactly solved \cite{re:BaxterBook}.
It is governed by the parameter \cite{re:BaxterBook}
\begin{equation} \label{Delta}
\Delta \ident {{a^2+b^2-c^2}\over{2ab}}
       = {{\sinh {(-2\lambda \hat\beta)}}\over{\sinh {2\hat\beta}}}
       \stackrel{\hat\beta\goesto 0}{\longrightarrow}\; -\lambda .
\end{equation}
Here we use periodic boundary conditions for
the vertex lattice.
The normal XXZ boundary conditions would be periodic in
both the spin chain direction
and the Trotter direction, and correspond to
periodic boundary conditions
in the 45-degree directions for the vertex lattice.

Monte Carlo simulations of constrained systems like the 6-vertex model
have suffered from severe critical slowing down in the past.
The situation has been drastically improved by the
advent of the loop cluster algorithm \cite {re:CSD_V1,re:CSD_V2,re:CSD_V4},
which we now sketch very briefly.
In terms of arrows, the constraint of Eq.~(\ref{continuity}) is a
``divergence = zero'' condition.
All arrows therefore lie on directed closed loops (like magnetic field
lines).
The loop algorithm stochastically constructs such a loop and
flips the direction of all arrows on the loop
in a single Monte Carlo step while maintaining detailed balance.
It can therefore perform large steps in phase space,
and produce statistically independent configurations within a few
Monte Carlo steps even at infinite correlation length on large
lattices \cite {re:CSD_V1}.
Note that the loop algorithm also easily produces
loops that wind around the lattice.
At the KT transition $({a\over c} = {b\over c} = {1\over2})$, for example,
autocorrelation times are reduced by
about two orders of magnitude on a $64^2$ lattice with respect to
a Metropolis simulation.

\section{MCRG Method}

Here we use the MCRG method described in detail in Ref.\ \cite{re:MunNov}.
The idea of this MCRG method is to let the Monte Carlo simulation itself
provide the renormalization group to be used.  This is done by building
on the MCRG method developed by Swendsen \cite{re:RHS1,re:RHS2} to calculate
critical exponents.  Rather than use a block spin procedure in real space,
the RG is defined in momentum space in a similar fashion as the procedure
introduced by Swendsen in Ref.\ \cite{re:RHS3}.
Rather than using the Fourier transform of the configurations
\cite{re:RHS3}, we use the Fourier transform of each state
individually \cite{re:Rikvold1},
\begin{equation}
\label{eq:fft}
\hat{v}_{\bf k}^{\alpha} =
\sum_{\bf r}\delta_{\alpha,v_{\bf r}}\exp{(i{\bf k}\cdot{\bf r})},
\end{equation}
where $\alpha$ labels the vertex state $v$, and the summation is over the
simulated lattice.  In our case, $\alpha$ takes the values of one through
six.  An inverse Fourier transform over a restricted part of the
momentum-space is then performed.  The same restricted part is used for all
generated configurations, and the inverse transform is given by
\begin{equation}
\label{eq:ifft}
v^{\prime}_{\bf s}{}^{\alpha} =
             \sum_{\bf k}{}^{\prime}\hat{v}_{\bf k}^{\alpha}
   \exp{\bigg(-i({\bf k}-{\bf m}_{\alpha})\cdot{\bf s}\bigg)},
\end{equation}
where ${\bf m}_{\alpha}$ is a shift in momentum space, and the prime on the
summation indicates that the sum is over a restricted part of momentum
space.  Finally, on each site of the reduced lattice for each configuration
the state $\alpha$ is chosen for which the real part,
$\Re(v^{\prime}_{\bf s}{}^{\alpha})$ is largest.
Of course this choice is arbitrary, since for example the state with the
largest modulus could have been chosen.

Both the restricted part of momentum space and the shift ${\bf m}_{\alpha}$
are the same for all generated configurations and are
given by the Monte Carlo simulation itself.  The shift ${\bf m}_{\alpha}$
is chosen such that the important fluctuations, as determined by the
peak in the momentum-space plot from the entire Monte Carlo simulation,
are shifted to the center of the Brillouin zone.  This transforms each
state into a mainly \lq ferromagnetic' state, and allows the use of normal
ferromagnetic operators to obtain critical exponents from the
linearized transformation matrix ${\bf T}^*$.
The restricted region of
the shifted momentum space is chosen to be the region in the center of
the Brillouin zone that gives $N^\prime$$=$$L^\prime$$\times$$L^\prime$
spins on transforming back to real space.  Note that this MCRG procedure
does not preserve the constraints of the vertex model (the continuity
of loops).  However, the renormalized variables should be considered to be
composite variables, and the vertex constraints on the original lattice
are taken into account during the RG procedure in a way dictated by the
Monte Carlo simulation of the model.

In principle, one could avoid transforming back to real space, but at
the expense of defining the \lq majority rule' for the states in
momentum space, and at the expense
of defining operators for the linearized transformation
matrix ${\bf T}^*$ in momentum space; neither of which has been attempted
to the authors' knowledge.  The transformation back to real space reduces
the lattice size and gives the RG.  However, a normalization is also
required.  This is done by multiplying the new real-space spins with a
constant such that
$\sum|v_{\bf r}|^{2}/N =
\sum^{\prime} |v_{\bf s}^{\prime}|^{2}/N^{\prime}$,
where $N^{\prime}$ is the number of lattice sites on the reduced lattice,
and the primed summation is over renormalized spins $v_{\bf s}^{\prime}$
on the reduced lattice.  Other normalization methods can be
devised \cite{re:MunNov}, but they should all lead to the same critical
exponents from the MCRG.

The critical properties are obtained in the usual fashion from the
linearized RG transformation matrix ${\bf T}^*$ from the elements
\begin{equation}
\label{eq:RG1}
T^*_{\alpha\beta} = {{\partial K_\alpha^{(n)}}\over{\partial K_\beta^{(n+m)}}},
\end{equation}
where $K_\alpha^{(n)}$ is the coupling constant corresponding to the
operator $S_\alpha$ following the $n^{th}$ RG transformation.  The
$K_\alpha$ include factors of $1/k_B T$.  From the chain rule,
\begin{equation}
\label{eq:RG2}
{{\partial \langle S_\gamma^{(n)}\rangle}\over{\partial K_\beta^{(n+m)}}}
= \sum_\alpha T^*_{\alpha\beta}
{{\partial \langle S_\gamma^{(n)}\rangle}\over{\partial K_\alpha^{(n+m)}}}\;,
\end{equation}
where the angular brackets indicate the thermal average of the
operator.  The partial derivatives can be easily calculated from the
canonical ensemble:
\begin{equation}
\label{eq:RG3}
{{\partial \langle S_\gamma^{(n)}\rangle}\over{\partial K_\alpha^{(n+m)}}}
= \langle S_\gamma^{(n)} S_\alpha^{(n+m)} \rangle -
\langle S_\gamma^{(n)} \rangle \langle S_\alpha^{(n+m)} \rangle.
\end{equation}
The thermal averages in Eq.~(\ref{eq:RG3}) are calculated between the
renormalized lattices after $n$ and $n$$+$$m$ RG transformations.
These averages can be calculated at the simulated set of parameters, but
they can also be calculated with parameters other than the simulated
ones using a single reweighting method
\cite{re:RW1,re:RW2,re:RW3,re:RW4}, or using data from multiple
simulations for different parameters \cite{re:RW4,re:RW5,re:MunNov}.

The number of operators in Eq.~(\ref{eq:RG2}) is infinite.  However, the
MCRG estimate for the matrix ${\bf T}^*$ is found from Eq.~(\ref{eq:RG2})
by first truncating all the matrices in Eq.~(\ref{eq:RG2}) and then
multiplying Eq.~(\ref{eq:RG2}) by the inverse of the truncated matrix
that multiplies ${\bf T}^*$ in Eq.~(\ref{eq:RG2}).  To first order in
perturbation theory, it has been shown that this procedure takes into
account the operators which have not been included in the truncation
procedure \cite{re:Shankar}.
Since we have performed the shift ${\bf m}_{\alpha}$ for
each state, we need to include only \lq ferromagnetic' operators.

Once the largest eigenvalues, $\lambda_1$$\ge$$\lambda_2$$\ge$$\cdots$,
of the linearized transformation matrix ${\bf T}^*_{\alpha\beta}$ are
determined, the eigenvalues of the RG are given by
\begin{equation}
\label{eq:yRG1}
y_i = \ln(\lambda_i)/\ln(b),
\end{equation}
where the scale factor $b$ is given by the ratio of the lengths on the
two lattice sizes used,
$b$$=$$L_{large}^\prime$$/$$L_{small}^\prime$$=$$(N_{large}^\prime$$/$
$N_{small}^\prime$$)^{1/d}$,
where $d$ is the dimension of the model studied.
The exponents $y_i$ are irrelevant if they are less than zero, are
marginal if $y_i$$=$$0$, and are relevant and give universal critical
exponents for models associated with the fixed point if $y_i$$>$$0$.

For a second-order transition the two largest critical exponents
of the RG ($y_T$ from the operators in the thermal section and $y_H$ from
operators in the magnetic section) allow the critical exponents to be
calculated in the
normal fashion from $\nu$$=$$1$$/$$y_T$ and
$\eta$$=$$2$$+$$d$$-$$2y_H$.

Special care must be taken in the case of a model with a
Kosterlitz-Thouless (KT) transition\cite{re:KTMCRG}.  The authors in
Ref.~\cite{re:KTMCRG} show that although the magnetic exponents are given by
$\eta$$=$$2$$+$$d$$-$$2y_H$ in the normal fashion, the values of
$\nu$ for a KT transition must be obtained by fitting to the KT RG
equations.  We have not done such a fitting procedure, but have rather
concentrated
only on the magnetic exponents.  In general one can also use the MCRG
method to obtain the values of the critical
couplings\cite{re:KTMCRG,re:Rajan}.
However since the critical couplings for the $d$$=$$1$
XXZ model are known exactly, we have not tried to find them using MCRG.

\begin{figure}[tbp]
\vspace*{.6cm}
\center{\em (See \cite{re:BaxterBook}, fig.\ 8.5)}
\vspace*{.6cm}
\caption{The phase diagram for the 6-vertex model is
shown.
The vertex weights satisfy $w_1$$=$$w_2$,
$w_3$$=$$w_4$,
and $w_5$$=$$w_6$.
The dashed line corresponds to the F-model where
$w_1$$=$$w_3$ as well, and this line terminates at the `infinite temperature'
point where $w_1$$=$$w_5$ (filled circle).  The open circles represent the
points where we have made MCRG calculations.
\label{fig2}}
\end{figure}

\section{Theoretical results for the $d$$=$$1$ XXZ Model}

The $d$$=$$1$ spin-${1\over 2}$ XYZ model has been studied
 using a wide variety of methods.  These methods have
ranged
from exact methods\cite{re:BaxterBook,re:EXACT1,re:EXACT2},
to universality and mapping methods\cite{re:denNijs,re:BlackEmery},
to conformal invariance methods\cite{re:CARDYbook,re:ABBci},
to series expansion methods\cite{re:series1},
to exact diagonalization
studies\cite{re:BonnerFisher,re:Moreo1,re:Moreo2},
to Monte Carlo methods\cite{re:Suzuki3,re:CullLan,re:Lyklema}.
The phase diagram which has emerged from these studies is illustrated in
the 6-vertex representation in Fig.~2\cite{re:BaxterBook}.
We have performed our MCRG calculations only on the F-model (where
$w_1$$=$$w_2$$=$$w_3$$=$$w_4$ and $w_5$$=$$w_6$) at points
illustrated by the open circles in Fig.~2.
In Fig.~2 region~I corresponds to a ferroelectric region
where the lowest energy
state is one with all vertices equal to 1 or~2.  In this region the
excited states give a negligible contribution to the partition function
and the system is frozen in one of the two ground states.
The situation is similar for
region~II, where all vertices equal 3 or~4, and region IV where the two
ordered states have vertices of 5 on one sublattice and 6 on the other
sublattice.
Region III is the disordered phase where there is no
spontaneous order, nor interfacial tension, but the correlation length
is infinite and correlations decay as an inverse power of distance
rather than exponentially\cite{re:BaxterBook}.

The F-model with $w_1$$<$$w_5/2$ (which corresponds to the asymmetry
parameter of Eq.~(\ref{eq:XXZHam}), $\lambda$$>$$1$)
 is in region~IV.
Since in this region the model is completely ordered,
one expects to obtain a low-temperature discontinuity fixed point
(the Nienhuis-Nauenberg criterion\cite{re:NienNauen})
with the largest RG eigenvalue $y$$=$$d$.

There is a
Kosterlitz-Thouless (KT) transition in the F-model when $w_1$$=$$w_5/2$
(where $\lambda$$=$$1$, and the model is the Heisenberg model).
In the XXZ model there is a KT transition from the massless into the
antiferroelectric ordered domain at $\lambda$$=$$1$.  This
occurs through a Kosterlitz-Thouless mechanism caused by the excitations of
an umklapp process\cite{re:denNijs,re:BlackEmery}.  At $\lambda$$=$$1$
the vortex operator (which comes from the critical exponent $x_T^{8V}$ of
the 8-vertex model) is marginal.

For the F-model in region~III the critical exponents are those of the
critical 8-vertex model\cite{re:BaxterBook} and are the same as those
of the Luttinger model\cite{re:denNijs}.  From conformal
invariance\cite{re:ABBci} the entire operator content is known.
Stringlike solutions of the Bethe-{\it Ansatz\/} equations yield excited
states corresponding to operators $O_{i,j}$ with dimensions
\begin{equation}
\label{eq:CI1}
x_{i,j} = i^2 x_p + j^2/4x_p \quad \quad i,j=0,1,2,\cdots.
\end{equation}
These operators are the analogs of the Gaussian-model operators composed of
spin-wave excitations of index $i$ and a ``vortex" of vorticity $j$.
In Eq.~(\ref{eq:CI1}) $x_p$$=$$($$\pi$$-$$\gamma$$)$$/$$2$$\pi$,
where $\gamma$$=$${\rm arccos}$$($$\lambda$$)$.  
The operators $O_{1,0}$ and $O_{2,0}$ correspond respectively to the
polarization and energy operators of the 8-vertex model.
The operator $O_{0,1}$ is the crossover operator of the 8-vertex model, or
equivalently the energy operator of the Ashkin-Teller model.
{}From the dimensions $x_{i,j}$ the RG exponents are found through
\begin{equation}
\label{eq:CI2}
y_{i,j} = d - x_{i,j} .
\end{equation}
The exponents $\eta_{i,j}$$=$$2$$+$$d$$-$$2y_{i,j}$$=$$2$$x_{i,j}$
govern the behavior of correlations.  For example, for $O_{0,1}$ one has
$\eta_{0,1}$$=$$2$$x_{0,1}$$=$$2$$+$$d$$-$$2$$y_{0,1}$ and this critical
exponent governs the correlation $\eta_z$$=$$\eta_{0,1}$ with
\begin{equation}
\label{eq:Cz}
\left\langle S^z(0) S^z(r) \right\rangle \sim r^{-\eta_z}
\end{equation}
where $S^z$$=$$1\over2$$\sigma^z$.

\section{MCRG results}

\begin{figure}[tbp]
\vspace*{.6cm}
\center{\em (See \cite{re:MunNov}, fig.\ 4)}
\vspace*{.6cm}
\caption{The 11 operators used for the truncated linearized RG
transformation matrix ${\bf T}^*$ are shown.
Filled circles represent a vertex in any state, circles with numbers
inside represent a $\delta$-function for that vertex state, and the
\lq bull's eye' in the last operator is any vertex which is not a 5 or~6.
The square lattice is drawn with light lines, while heavy lines join the
interacting sites.  For the operator to be nonzero all filled-circle
interacting vertices must have the same vertex state.  All combinations
which are related by translation, rotation, and reflection to those shown
were included in the calculation.
\label{fig3}}
\end{figure}

We have applied the MCRG method described in Sec.~2 to the 6-vertex model
on $L$$\times$$L$ lattices with periodic boundary conditions.
The 11~operators which we included in the truncated linearized RG
transformation
matrix ${\bf T}^*$ are shown in Fig.~\ref{fig3}.  Note that because
the momentum-space RG does not conserve the 6-vertex constraint
on the renormalized lattice, we have not included polarization-type
operators in the 11 measured operators.  Our calculations were done on
a CRAY Y-MP432 supercomputer, and the Fourier transforms used were
FFT's in the NAG library (the explicit routines used depended on the
lattice sizes the FFT was applied to).

We first studied the F-model with $w_1$$=$$w_5$$/$$4$ in the ordered
region~IV of Fig.~2, and the results are
shown in Table~\ref{table1}.  The lattice simulated for Table~\ref{table1}
was $32$$\times$$32$ and the renormalized lattices were
$L^\prime$$\times$$L^\prime$ with $L^\prime$$=$$27$, 25, 21, 17, and 15.
The RG analysis was performed using $15000$ configurations
which were generated skipping 250 Monte Carlo cluster updates between
configurations and with $250000$
cluster updates for thermalization.
The exponential autocorrelation time, $\tau_{\rm exp}$ in units of
`sweeps'\cite{re:CSD_V4} is given by the autocorrelation time in units of
cluster updates times the average cluster size divided by $2$$L^2$.
Here $\tau_{\rm exp}$$=$$43(5)$, and it
is most clearly visible in the sublattice energy\cite{re:CSD_V4}.
The average cluster size is 0.96 times $2$$L^2$.
In principle one can obtain estimates for the statistical errors of the
eigenvalues in Table~\ref{table1}, for example by analyzing $J$ bins of
the generated configurations.  We have not yet performed such an analysis.

\begin{table}[tb]

\centering
\begin{tabular}{|c|c@{}cccc|}
\hline
\hline
\multicolumn{1}{|c|}{$n$} &$m$& 1 & 2 & 3 & 4 \\
\hline
  & & \multicolumn{4}{c|}{Largest RG Exponent} \\
1 & & 1.9998 & 1.9996 & 1.9981 & 1.9864 \\
2 & & 2.0003 & 1.9980 & 1.9868 &        \\
3 & & 1.9963 & 1.9840 &        &        \\
4 & & 1.9811 &        &        &        \\
\hline
  & & \multicolumn{4}{c|}{Second RG exponent} \\
1 & & 0.6449 & 1.0041 & 1.2214 & 1.1063 \\
2 & & 0.7962 & 1.1325 & 1.0128 &        \\
3 & & 0.9313 & 0.7794 &        &        \\
4 & & 0.9683 &        &        &        \\
\hline
\hline
  & & \multicolumn{4}{c|}{RG scale factor $b$} \\
1 & & 1.080 & 1.286 & 1.588 & 1.800 \\
2 & & 1.191 & 1.471 & 1.667 &       \\
3 & & 1.235 & 1.400 &       &       \\
4 & & 1.133 &       &       &       \\
\hline
\hline
\end{tabular}
\caption{The two largest exponents of the linearized transformation
matrix, ${\bf T}^*$, are listed for the simulation in the
ordered phase at $w_1/w_5$$=$${1\over 4}$.  The lattice simulated was a
$32$$\times$$32$ lattice, and the sizes of the renormalized
$L'$$\times$$L'$ lattices were $L'$$=$27, 25, 21, 17, and 15.
Listed are the exponents between renormalized lattices $n$ and $n$$+$$m$ as
well as the corresponding RG scale factors $b$.  For example, the entry with
$n$$=$$2$ and $m$$=$$3$ corresponds to an RG transformation between lattices
with $L'$$=$$25$ and $L'$$=$$15$.
The data should converge for large $n,m$,
but also become susceptible to finite size effects.
The expected largest eigenvalue is
that of the zero-temperature fixed point since the system is
ordered in Region~IV of Figure~2.  Consequently the largest exponent
should be that of the discontinuity fixed point, $y$$=$$d$$=$$2$.
Our data converge to this value extremely rapidly.
The next largest eigenvalue would be the first analytic correction, which
approaches unity.
\label{table1}}
\end{table}

The values of Tables~1  and~2 should be read in a specific fashion to
see the convergence of the exponents.
In order to obtain good critical exponents from the MCRG one needs to be
able to penetrate the {\it linear\/} region about the fixed point, since
the exponents are calculated from the linearized transformation matrix
${\bf T}^*_{\alpha\beta}$.  It may take several iterations to
be able to get close enough to the fixed point that a linear approximation
is reasonable.  Consequently, the first iterations (the ones starting
from the largest lattices, and hence on the first lines in the tables)
may need to be ignored, or at least to be used only to see how the convergence
toward the linear region is proceeding.  For this reason we
do not show the iteration between the original lattice and the next
largest lattice in the tables.
The exponents should then converge for large values of $n$ and $m$.

If $n,m$ become too large, however, there are additional difficulties.
Smaller scale factors $b$
will have the larger statistical errors\cite{re:MunNov}.
Also, in MCRG studies the larger statistical errors typically occur on the
smallest lattices, which correspond to the last entry in each line of the
tables.  The smallest lattices will also have larger finite-size effects,
which are particularly disruptive for the longer-range operators included
in the calculation of ${\bf T}^*_{\alpha\beta}$.

In Table~\ref{table1} one sees extremely rapid
convergence to the eigenvalue $y$$=$$d$$=$$2$ associated with
the Nienhuis-Nauenberg criterion for a discontinuity fixed
point \cite{re:NienNauen}.
The next-largest eigenvalue, while not having converged too well, seems
to be approaching the value of unity, which would give the first
analytical correction to the discontinuity fixed-point behavior.

\begin{table}[tbp]

\centering
\begin{tabular}{|c|ccccc|||c|ccccc|}
\hline
\hline
\hline
  & & \multicolumn{4}{c|||}{$w_1/w_5=1/2$ ~~~ $\lambda=1$} &
  & & \multicolumn{4}{c|}{$w_1/w_5=1/\sqrt{2}$ ~~~ $\lambda=0$} \\
\hline
\hline
\multicolumn{1}{|c|}{$n$} &$m$& 1 & 2 & 3 & 4 &
\multicolumn{1}{|c|}{$n$} &$m$& 1 & 2 & 3 & 4 \\
\hline
  & & \multicolumn{4}{c|||}{Largest RG Exponent} &
  & & \multicolumn{4}{c|}{Largest RG Exponent} \\
1 & & 2.022 & 1.809 & 1.707 & 1.667 & 1 & & 1.902 & 1.477 & 1.342 & 1.299 \\
2 & & 1.676 & 1.624 & 1.605 &       & 2 & & 1.098 & 1.145 & 1.165 &       \\
3 & & 1.591 & 1.581 &       &       & 3 & & 0.975 & 1.070 &       &       \\
4 & & 1.563 &       &       &       & 4 & & 0.844 &       &       &       \\
\hline
  & & \multicolumn{4}{c|||}{Second RG Exponent} &
  & & \multicolumn{4}{c|}{Second RG Exponent} \\
1 & & 1.466 & 1.068 & 0.835 & 0.763 & 1 & & 1.185 & 0.493 & 0.175 & 0.198 \\
2 & & 0.636 & 0.585 & 0.576 &       & 2 & & $-$.520 & $-$.234 & $-$.005 &   \\
3 & & 0.436 & 0.500 &       &       & 3 & & $-$.610 & $-$.166 &       &    \\
4 & & 0.358 &       &       &       & 4 & & $-$.931 &       &       &       \\
\hline
\hline
\hline
\end{tabular}
\vskip 0.05 true cm
\centering
\begin{tabular}{|c|ccccc|}
\hline
\hline
\hline
  & & \multicolumn{4}{c|}{$w_1/w_5=0.5825\cdots$ ~~~ $\lambda=1/2$} \\
\hline
\hline
\multicolumn{1}{|c|}{$n$} &$m$& 1 & 2 & 3 & 4  \\
\hline
  & & \multicolumn{4}{c|}{Largest RG Exponent} \\
1 & & 2.050 & 1.715 & 1.581 & 1.524 \\
2 & & 1.501 & 1.463 & 1.437 &       \\
3 & & 1.397 & 1.391 &       &       \\
4 & & 1.305 &       &       &       \\
\hline
  & & \multicolumn{4}{c|}{Second RG Exponent}  \\
1 & & 1.429 & 0.908 & 0.616 & 0.537 \\
2 & & 0.292 & 0.307 & 0.327 &       \\
3 & & 0.029 & 0.216 &       &       \\
4 & & $-$.137 &       &       &       \\
\hline
\hline
\multicolumn{1}{|c|}{$n$} &$m$& 1 & 2 & 3 & 4  \\
\hline
  & & \multicolumn{4}{c|}{RG scale factor $b$} \\
1 & & 1.244 & 1.645 & 2.429 & 3.400 \\
2 & & 1.323 & 1.952 & 2.733 &       \\
3 & & 1.476 & 2.067 &       &       \\
4 & & 1.400 &       &       &       \\
\hline
\hline
\hline
\end{tabular}
\caption{The two largest exponents of the linearized transformation
matrix ${\bf T}^*$ are listed for the simulations
at $\lambda$$=$$0$, $1\over2$, and $1$.
The lattice simulated was
$64$$\times$$64$, and the sizes of the renormalized
$L'$$\times$$L'$ lattices were $L'$$=$51, 41, 31, 21, and 15.
Listed are the exponents between renormalized lattices, $n$ and $n$$+$$m$, as
well as the corresponding RG scale factors $b$.  For example, the entry with
$n$$=$$3$ and $m$$=$$2$ corresponds to an RG transformation between lattices
with $L'$$=$$31$ and $L'$$=$$15$.
The data should converge for large $n,m$;
but  should also become susceptible to finite size effects.
For $\lambda$$=$$0$, ${1\over 2}$, and $1$, the largest eigenvalues
$y_z$$=$$2$$-$$\eta_z$, corresponding to the operator $O_{0,1}$, are
$1$, ${5\over 4}$ and ${3\over 2}$, respectively.
\label{table2}}
\end{table}

Our other MCRG results are shown in Table~\ref{table2}.  These are at the
KT point ($\lambda$$=$$1$), the XY point (free-fermion case with
$\lambda$$=$$1$$/$$\sqrt{2}$), and an intermediate point with
$\lambda$$=$$1$$/$$2$.  Table~\ref{table2} lists the two largest
eigenvalues obtained from our MCRG procedure.
Each value of $\lambda$ studied took approximately 23 hours of Y-MP time
for the RG analysis of the generated configurations.  The time required
for the Monte Carlo portion of the runs depended on the value of
$\lambda$ since the cluster-size per volume depended on $\lambda$.
For $\lambda$=0, ${1\over 2}$, and 1 the cluster-size
was $0.15$, $0.09$, $0.038$ times $2$$L^2$
and the Y-MP CPU time used corresponded to
3.9, 2.5, and 1.4 hours.
The lattice simulated for Table~\ref{table2}
was $64$$\times$$64$, and the renormalized lattices were
$L^\prime$$\times$$L^\prime$ with $L^\prime$$=$$51$, 41, 31, 21, and 15.
The RG analysis was again performed using $15000$ configurations
which were generated skipping 250 cluster updates between
configurations and using $250000$
cluster updates for thermalization. Consequently, a total of
4$\times$10$^6$ cluster updates were generated.
The exponential autocorrelation time, which is most clearly
visible in the sublattice energy\cite{re:CSD_V4}, for $\lambda$=0,
${1\over 2}$, and 1, is
$\tau_{\rm exp}$$=$$4.0(2)$, $8.1(9)$, and $5.7(6)$, respectively.

 The convergence toward the linear region about the fixed point
is expected to be particularly slow when a marginal eigenvalue is present,
as is the case in Table~2.  Consequently, in Table~2 convergence has not yet
occured, but the trend toward convergence of the largest eigenvalue is
evident and in agreement with the expected result.

In Table~\ref{table2} we see only one nearly converged
largest exponent $y$ for
each value of $\lambda$.  This exponent can be associated with the
operator $O_{0,1}$ corresponding to the cross-over exponent of the
8-vertex model, and the exponent $\eta_z$ can be associated with the
$S^z$ correlations.  The reason operators such as $O_{1,0}$ may not be
found in our MCRG calculation is because such operators correspond to
the polarization operator of the 8-vertex model, and such asymmetric
operators are not included in the operators shown in Fig.~3.
Such operators are not easily included since the momentum-space RG
procedure does not preserve the local vertex constraint.

\section{Discussion and Conclusions}

We have implemented for the first time the marriage of a loop algorithm
to alleviate critical slowing down and an MCRG procedure that does not
necessitate that one know the symmetries of the underlying classical
equivalent beforehand since they are given by the Monte Carlo.
We have applied the resulting MCRG method to the study of the $d$$=$$1$
spin-$1\over 2$ quantum XXZ chain.  Although this model has a
Kosterlitz-Thouless transition, and consequently the convergence toward the
fixed point is fairly slow, we are able to obtain the
critical exponent $\eta_z$ associated with the $S^z$ correlation function
reasonably well.  The exponent $\eta_z$ comes from
the operator $O_{0,1}$, which is associated with the crossover exponent of the
8-vertex model.  We have pointed out the difficulties in our study that
have prevented us from obtaining the other exponents of the XXZ model.
Although this is a preliminary study, our MCRG method
shows a great deal of promise, since it is easily generalized
to the study of many other quantum
models in one and higher dimensions.

\vskip 0.4 true cm
\pagebreak[4]
\noindent{\bf ~~~~Acknowledgments}

The authors wish to thank J.~Adler, M.~Marcu, A.~Moreo, and P.~A. Rikvold
for useful discussions.
The MCRG program used was based on code partially written by E.~P.~M\"unger.
This research was supported in part by the Florida State University
Supercomputer Computations Research Institute which is partially funded
through contract \# DE-FC05-85ER25000 by the U.S. Department
of Energy.

\end{document}